\documentstyle[aps,prl,epsf]{revtex}
\begin{document}
\draft
\title{Spin waves in a two dimensional $p$-wave 
superconductor: Sr$_2$RuO$_4$}

\author{Hae-Young Kee$^{1}$, Yong Baek Kim$^{2}$, and Kazumi Maki$^{3}$}
\address{
$^1$ Department of Physics, University of California, 
Los Angeles, CA 90095\\
$^2$ Department of Physics, The Ohio State University, 
Columbus, OH 43210\\
$^3$ Department of Physics, 
University of Southern California, Los Angeles, CA 90089}

\date{\today}
\maketitle

\begin{abstract}
We study spin excitations in a two dimensional 
$p$-wave superconductor with ${\vec \Delta} = 
{\hat d}({\hat k}_1 \pm i {\hat k}_2)$ symmetry in the context of 
the newly discovered superconducting Sr$_2$RuO$_4$. 
The polarization and spectrum of spin wave excitations 
are identified and their experimental consequences 
are discussed. 
\end{abstract}


\section{Introduction}

Recent discovery of superconducting Sr$_2$RuO$_4$ generates
a lot of efforts to determine the pairing symmetry of a possible
unconventional superconducting order parameter in this system.
Being a 4$d$-orbital analogue\cite{cava} of the high $T_c$ 
cuprate superconductors, Sr$_2$RuO$_4$ has the same layered perovskite 
structure as La$_2$CuO$_4$ and becomes superconducting below 
$T_c \approx 1.5K$\cite{maeno1,maeno1-1,maeno2}.
Despite the structural similarity, 
it behaves very differently from the copper oxides.
The normal state below $50K$ can be well described by a quasi-2D 
Landau-Fermi liquid: The resistivity shows $T^2$ behavior in the 
$a$-$b$ plane and $c$-direction with a large anisotropy ratio\cite{maeno2}. 
Quantum oscillations\cite{mac1} revealed three 
cylindrical Fermi surface sheets in accordance with
the band structure calculations\cite{band}. There are three 4$d$-orbitals
($d_{xy},d_{yz},d_{zx}$) of the Ru$^{4+}$-ions which form
three bands crossing the Fermi level. There are two electron-like
and one hole-like Fermi surfaces\cite{mac1}.
On the other hand, the mass enhancement ($\approx 4$)
in this material is not small indicating the existence of
strong correlation\cite{mac1}.
The facts that it has a large mass enhancement and a related
material, SrRuO$_3$, shows ferromagnetism lead to the 
proposal that the superconducting state of Sr$_2$RuO$_4$
is formed by odd-parity pairing (spin-triplet pairing)
presumably in the $p$-wave channel\cite{sigrist1,baskaran,mazin2}.

There have been a number of experiments indicating
unconventional superconductivity in Sr$_2$RuO$_4$.
The transition temperature was found to be very sensitive
to non-magnetic impurities\cite{mac2} and nuclear 
quadrupole resonance
(NQR) found no Hebel-Slichter peak\cite{maeno4}.
Even though these experiments\cite{mac2,maeno4,liu} suggested 
that the pairing symmetry is non-s-wave, they could not determine 
the pairing symmetry itself.
More recently $^{17}$O-Knight shift in NMR was measured and it 
is consistent with the spin triplet superconductivity\cite{ishida}
with ${\hat d}$ parallel to the $c$ axis as the case of superfluid 
$^3$He -A\cite{vollhardt}.
Here ${\hat d}$ is the unit vector of the triplet order 
parameter\cite{vollhardt}.
On the other hand, the specific heat data are consistent with 
usual s-wave superconductors if we subtract a persistent $T$-linear 
term\cite{nishizaki,nishizaki2}.
In the earlier experiment the ratio of this coefficient of 
the $T$-linear term to the one in the normal state 
$\gamma_0/\gamma_N$ was larger than 0.5.
This initiated the proposal of possible nonunitary state in Sr$_2$RuO$_4$
\cite{sigrist2}.
However more recent data\cite{nishizaki2} show that 
$\gamma_0/\gamma_N < 0.25$.
This means the nonunitary state is untenable.
The specific heat data further imply that the energy gap should be 
almost independent of ${\bf k}$ like in an s-wave superconductor 
and also there should be a normal-state-like component.
These features are most naturally described by the three orbital band 
model\cite{agt1}, where it is assumed that the superconductivity resides
mainly in the $\gamma$ band, while the $\alpha$ and $\beta$ bands may be 
considered in the normal state.
Of course we cannot exclude the possible small superconducting order 
parameter associated with the $\alpha$ and $\beta$ band\cite{agt2}.
In this perspective, the small magnetization seen by the muon spin resonance
\cite{maeno5} is rather puzzling.
In addition, rather strong flux pinning in the vortex state is observed 
in Sr$_2$RuO$_4$ comparable to the one in $B$-phase of UPt$_3$ and 
U$_{0.97}$Th$_{0.03}$Be$_{13}$\cite{mota}.
The latter two systems are considered to be in the nonunitary state\cite{heffner}.
Here it is worthwhile to mention that, although all these are consistent 
with the triplet pairing, both the
magnetization and the strong vortex pinning should arise from 
some topological defects or dislocation in the samples.
A recent small angle neutron scattering experiment in a magnetic field
parallel to the $c$ axis showed that the square vortex lattice is
almost everywhere in the $B$-$T$ phase diagram\cite{mac3}.
This square vortex lattice is also very consistent with $p$-wave 
superconductivity we consider here\cite{agt3,agt4,wang}.

In the following we take the order parameter
\begin{equation}
{\vec \Delta}({\bf k})= \Delta {\hat d} 
({\hat k}_1\pm i {\hat k}_2) \ ,
\label{order}
\end{equation}
where ${\hat d}$ is assumed to be parallel to the $c$ axis 
and ${\bf k}$ is the quasi-particle wave vector within the $a$-$b$ plane.
For simplicity, we also assume that the superconductivity resides only in 
the $\gamma$ band.
This means the $\alpha$ and $\beta$ bands provide 
non-superconducting background.
Recently starting from Eq.\ref{order}, microscopic studies of the vortex 
state\cite{agt4,wang}, the effect of impurities\cite{maki},
and the quasi-particle spectrum around a single vortex\cite{matsumoto}
have been discussed.

Here we study spin dynamics in the $p$-wave superconductors with 
the order parameter given by Eq.\ref{order}.
It is important to realize that there are four distinct
situations for the measurement of the dynamical spin
susceptibility in two-dimensional $p$-wave superconductors.
This is because the direction of the vectorial order parameter
${\bf d}({\bf k})$ is fixed along a crystallographic axis
which is perpendicular to the basal plane.
One can apply a magnetic field along the directions
perpendicular and parallel to the RuO$_2$ plane.
In each case, one can measure longitudinal and
transverse susceptibilities. Therefore there exist
four different susceptibilities. This is a unique property
of the two dimensional $p$-wave superconductor because,
in three dimensional case, the direction of the order
parameter is not fixed and is always perpendicular to
the applied magnetic field.

We also assume that the pinning of ${\hat d}$ vector parallel to the $c$ axis 
involves a finite pinning energy $-{1 \over 2} \chi_N \Omega_d^2 (d_z)^2$,
most likely due to the spin-orbit coupling.
Here we do  not evaluate the exact value of $\Omega_d$, but
we expect that it will be $\Omega_d(T) < \Delta(T)$.
Recently the pinning frequency was estimated by
Tewordt.\cite{tewordt}

We believe that the experimental determination of $\Omega_d(T)$  will
provide an insight in the pinning mechanism of ${\hat d}$.
In order to avoid future misunderstanding, we shall first explain the 
role of ${\hat d}$ here.
${\hat d}$ is called the spin vector which has been used in studying
$^3$He.
It is perpendicular to the direction of the spin associated with the condensed 
pair\cite{vollhardt}.
Under these assumptions, we found the following results
for the spin excitations.
\vskip 0.3cm
\noindent
{\bf Case A}: When magnetic field is parallel to the $a$-$b$ plane
(i.e., ${\bf H} \perp {\hat d}$) 

We have the spin waves with the dispersion relations,
\begin{equation}
\omega_{\parallel}^2= (1-I f_d) \Omega_d^2 +\frac{1}{2} 
f_d (v_F q)^2
\label{mode}
\end{equation}
and
\begin{equation}
\omega_{\perp}^2=\omega_{\parallel}^2+\omega_L^2
\label{modet}
\end{equation}
for the longitudinal ($\omega_{\parallel}$) and the 
transverse ($\omega_{\perp}$) resonance, respectively.
The ``pinning'' frequency, $\Omega_d$, is associated
with the restoring tendency of the order parameter to
the initial direction ${\hat d}$ against external
perturbations.
Here $\omega_L=\mu_B H$ is the Larmor frequency.
$I(=-{1 \over 4} Z_0)$ is the dimensionless on-site Hubbard potential and
$f_d$ is the dynamical superfluid density ($\omega \gg v_F q$) given by
\begin{equation}
f_d=
\int_{\Delta}^{\infty} dE \frac{\Delta^2}{E^2} \frac{\tanh(\frac{E}{2T})}
{\sqrt{E^2-\Delta^2}},
\end{equation}
where $\Delta$ is the magnitude of the full gap.
These can be readily generalized in the presence of Fermi liquid 
corrections\cite{vollhardt}.

\vskip 0.3cm
\noindent
{\bf Case B}: When magnetic field is perpendicular to the $a$-$b$ plane
(i.e. ${\bf H} \parallel {\hat d}$)

The static susceptibility in this case is given by
\begin{equation}
\chi=\chi_0 \frac{Y}{1-I Y} \ ,
\end{equation}
where $Y=1-f_s$ is the Yoshida function\cite{vollhardt}.

There is no longitudinal collective mode, but a damped mode
with the  following dispersion exists in
the longitudinal response.

\begin{equation}
\omega = i \frac{1}{\sqrt{2}} v_F |q| \sqrt{(1-f_d)I} \ .
\label{damp}
\end{equation}

For the transverse response, we have the spin wave with
the same dispersion relation as Eq.\ref{modet}.

\vskip 0.3cm

The remainder of this paper is organized as follows.
In section II, we show the calculation of the dynamical 
spin susceptibilities for four different cases mentioned 
above. We conclude in section III. Some of the technical
details are relegated to the appendix.

\section{Dynamical Spin Susceptibility and Spin Waves}

Here we use Green's function method of Ref.\cite{maki2}.
The single particle Green's function in the Nambu space is given by
\begin{equation}
G^{-1}(i\omega_n,{\bf k})=i \omega_n - \left [ \xi_{\bf k} 
+\frac{\Omega_L}{2}({\vec \sigma} \cdot {\hat h}) 
\right ] \rho_3 
- \Delta ({\hat k}\cdot {\vec \rho}) \sigma_1 \ ,
\label{green}
\end{equation}
where $\rho_i$ and $\sigma_i$ are Pauli matrices acting on
the particle-hole and spin spaces respectively,
${\hat h}$ is the unit vector parallel to the static magnetic field.
$\omega_n = (2n+1) \pi T$ is the fermionic Matsubara 
frequency, 
$\Omega_L=\omega_L/(1-I)$, $\xi_{\bf k}={{\bf k}^2 \over 2m}-\mu$, and
$\Delta$ is the magnitude of the superconducting order parameter.
%
%
Then the spin susceptibility is expressed as 
the auto-correlation function of spin operators.
For example, the irreducible 
spin-spin correlation function, $\chi^{00}_{ij}
\sim \langle [S_i,S_j] \rangle_0$,
can be computed from
\begin{equation}
\chi^{00}_{ij} (i\omega_{\nu},{\bf q}) = T \sum_n
\sum_{\bf p} {\rm Tr} [\alpha_i G({\bf p},\omega_n)
\alpha_j G({\bf p}-{\bf q},i\omega_n-i\omega_{\nu})] \ ,
\end{equation}
where $\alpha_1=\rho_3\sigma_1, \alpha_2=\sigma_2,
\alpha_3 = \rho_3 \sigma_3$ are the spin vertices
and come from the expression of spin density:
$S_i \sim {\rm Tr}[\Psi^{\dagger}\alpha_i\Psi]$.
$\omega_{\nu} = 2\nu \pi T$ is the bosonic Matsubara frequency.

In the fully renormalized spin susceptibilities, it is crucial to 
include the fluctuation of the superconducting
order parameter related to the rotation of the ${\hat d}$ vector.
Now let us look at the specific cases.

\subsection{Magnetic field is perpendicular to
the order parameter: ${\bf H} \ || \ {\hat {\bf x}}$}

\subsubsection{Longitudinal susceptibility}

Since the direction of the magnetic field is in 
${\hat {\bf x}}$-direction, the longitual susceptibility 
is given by 
$\chi_{xx} \sim \langle [S_x,S_x] \rangle$ and 
can be obtained from 
\begin{equation}
\chi_{xx} = {\chi^0_{xx} \over 1 - I \chi^0_{xx}} \ ,
\end{equation}
where $I$ is the exchange interaction and $\chi^0_{xx}$
is the susceptibility irreducible with respect to $I$.
$\chi^0_{xx}$ includes the coupling between the spin and
the order parameter, and can be written as
\begin{equation}
\chi^0_{xx} = \chi^{00}_{xx} + 
{V_{xx} g {\bar V}_{xx} \over 1 - g \Pi_{xx}} \ ,
\end{equation}
where $g$ is the strength of the interaction which is
responsible for the superconductivity.
Here $\chi^{00}_{xx} \sim \langle [S_x,S_x] \rangle_0$
is the bare susceptibility, 
$V_{xx} \sim \langle [S_x, \delta \Delta_x] \rangle_0,
{\bar V}_{xx} \sim \langle [\delta \Delta_x, S_x] \rangle_0$
represent the coupling between the spin density and
the order parameter fluctuation, and 
$\Pi_{xx} \sim \langle [\delta \Delta_x, \delta \Delta_x] \rangle_0$
the correlation between the order parameter fluctuations.
In Nambu's notation, 
$\delta \Delta_x 
\sim \Delta {\rm Tr}[\Psi^{\dagger} \alpha_1 \rho_1 \sigma_1 \Psi]$.
Notice that $\chi^0_{xx}$ consists of two parts - the quasiparticle
contribution and the contribution from the excitation of
the condensate. 
In the case of the longitudinal response,
the $\Omega_L$
term in Eq.\ref{green} should be dropped out from the final expression.
The details of the computation of each correlation function 
can be found in the appendix.
We obtain ($\zeta = {\bf v}_F \cdot {\bf q}$)
\begin{eqnarray}
\chi^{00}_{xx} (\omega, {\bf q}) &=& N(0) 
\langle \frac{\zeta^2-\omega^2 f}{\zeta^2-\omega^2} 
 \rangle \ , \cr
V_{xx} (\omega, {\bf q}) &=& N(0) 
\langle \frac{\omega}{2 \Delta} f \rangle  = 
 {\bar V}_{xx} (\omega, {\bf q})\ , \cr
\Pi_{xx} (\omega, {\bf q}) &=& g^{-1}
-N(0)\langle \frac{\zeta^2-\omega^2}{4 \Delta^2} f \rangle \ ,
\label{result1}
\end{eqnarray}
where 
\begin{equation}
f (\omega,\zeta) = 4 \Delta^2 (\zeta^2-\omega^2) \int_{\Delta}^{\infty} 
dE \frac{\tanh{(E/2T)}}{\sqrt{E^2-\Delta^2}} 
\frac{(\zeta^2-\omega^2)^2-4 E^2(\omega^2+\zeta^2)+4 \zeta^2 \Delta^2}
{[(\zeta^2-\omega^2)^2+4 E^2(\omega^2-\zeta^2)+4 \zeta^2 \Delta^2]^2-16
\omega^2 E^2 (\zeta^2-\omega^2)^2} \ .
\end{equation}
and $\langle F \rangle = \int_0^{2\pi} 
\frac{d\phi}{2\pi} F(\phi)$. From now on, 
we set the density of state, $N(0) \equiv 1$.
In all these analysis we neglect $\Omega_L=\omega_L/(1-I)$ 
for simplicity.

Since the orientation of the order parameter is initially
fixed along one of the crystallographic direction,
there should be an energy scale, $\Omega_d$, associated with the
tendency to restore the original direction against external
perturbations. This ``pinning'' frequency enters in the correlation
between fluctuations of order parameters and leads to
\begin{equation}
\Pi_{xx}(\omega, {\bf q}) = g^{-1}
- \langle \frac{\zeta^2-\omega^2 + \Omega^2_d}{4 \Delta^2} f \rangle \ .
\end{equation}
Thus we get the following expression for $\chi^0_{xx}$,
\begin{equation}
\chi_{xx}^0=\langle \frac{1}{\zeta^2-\omega^2} \left( \zeta^2
+\frac{\omega^2 \Omega_d^2 f}{\omega^2-\zeta^2-\Omega_d^2} \right)
\rangle
\label{longx}
\end{equation}
In order to find a well-defined excitation, we consider the limit
$\omega \gg v_F q$, where the quasiparticles do not generate
dissipation which could make collective modes damped.
In this case, the above expression can be
simplified in the long wavelength limit ($q \ll k_F$) as
\begin{equation}
\chi^0_{xx} \approx  \left [
-{1 \over 2} \frac{(v_F q)^2}{\omega^2} +
\frac{\omega^2  \Omega_d^2 f_d}{ ( {1 \over 2} (v_F q)^2 -\omega^2 )
(\omega^2-{1 \over 2} (v_F q)^2 -\Omega_d^2) }
\right ] \ .
\end{equation}
Finally, in the limit $\omega \gg v_F q$, we get the following
full susceptibility after taking into account the exchange interaction:
\begin{eqnarray}
\chi_{xx} &\approx& 
\frac{\omega^2 \Omega_d^2 f_d + {1 \over 2} (v_F q)^2 (\omega^2 -\Omega_d^2)}
{ ({1 \over 2} (v_F q)^2-\omega^2)(\omega^2-{1 \over 2} (v_F q)^2-\Omega_d^2)
-I[ {1 \over 2} (v_F q)^2 (\omega^2-\Omega_d^2)
+\omega^2 \Omega_d^2 f_d] } \,
\end{eqnarray}
where $f_d = {\rm lim}_{\omega \rightarrow 0}
{\rm lim}_{q \rightarrow 0}
\langle f \rangle$ and it is given by
\begin{equation}
f_d = \int_{\Delta}^{\infty} dE
{\tanh{(E/2T)} \over \sqrt{E^2-\Delta^2}} {\Delta^2 \over E^2} \ .
\end{equation}
It is called the ``dynamical superfluid density''.
Now we can read off the dispersion relation of a collective
mode from the pole of the response function.
We find that the longitudinal susceptibility supports
a spin wave and its dispersion relation is given by
\begin{equation}
\omega^2=(1-I f_d) \Omega_d^2+\frac{1}{2} f_d (v_F q)^2.
\label{mode2}
\end{equation}

This is consistent with the similar expression in superfluid
$^3$He-A phase\cite{vollhardt,note}.
Not surprisingly p-wave superconductors have the longitudinal spin
wave as in superfluid $^3$He-A.

\subsubsection{Transverse susceptibility}

Now we analyze the transverse susceptibility. 
The transverse susceptibility,
$\chi_{+-} \sim {1 \over 2} \langle [S_y+iS_z,S_y-iS_z]
\rangle$, can be obtained from
\begin{equation}
\chi_{+-} = { (1 - I \chi^0_{-+}) \chi^0_{+-} +
I \chi^0_{++} \chi^0_{--} \over
(1 - I \chi^0_{+-})(1 - I \chi^0_{-+}) -
I^2 \chi^0_{++} \chi^0_{--} } \ ,
\label{rpat}
\end{equation}
where $\chi^0_{(\pm,\pm)}$ are
the susceptibilities irreducible to $I$. One can easily show
that $\chi^0_{+-} = \chi^0_{-+}$ and $\chi^0_{++} = \chi^0_{--}$.
Now $\chi^0_{+-}$ and $\chi^0_{++}$ are given by
\begin{eqnarray}
\chi^0_{+-} &=& \chi^{00}_{+-} +
\sum_{k=y,z} {V_{+k} g {\bar V}_{k-} \over 1 - g \Pi_{kk}} \ , \cr
\chi^0_{++} &=& \chi^{00}_{++} +
\sum_{k=y,z} {V_{+k} g {\bar V}_{k+} \over 1 - g \Pi_{kk}} \ ,
\label{mixing}
\end{eqnarray}
Here $\chi^{00}_{+-} \sim \langle [S_y+iS_z,S_y-iS_z] \rangle_0$,
$\chi^{00}_{++} \sim \langle [S_y+iS_z,S_y+iS_z] \rangle_0$
are the bare susceptibilities.
$V_{\pm k} \sim \langle [S_y \pm i S_z,\delta \Delta_k] \rangle_0$,
${\bar V}_{k \pm} \sim \langle [\delta \Delta_k,S_y \pm i S_z] \rangle_0$
and $\Pi_{kk} \sim
\langle [\delta \Delta_k,\delta \Delta_k] \rangle_0$
are spin/order-parameter couplings and the fluctuation propagator
of the order parameters respectively, where
$\delta \Delta_k \sim {\rm Tr}[\Psi^{\dagger}
\alpha_k \rho_1 \sigma_1 \Psi]$.

When $\omega_L = 0$, we find
\begin{eqnarray}
\chi^{00}_{+-} &=& \chi^{00}_{-+} =
 \langle \frac{\zeta}{\zeta-\omega} \rangle
- \langle \frac{\zeta+\omega}{\zeta-\omega}\frac{f}{2}
\rangle \ , \ \ \
\chi^{00}_{++} = \chi^{00}_{--} =
  \langle \frac{f}{2} \rangle \ , \cr
V_{\pm y} &=&  {\bar V}_{y \pm} =  \frac{1}{\sqrt{2}}
\langle \frac{\omega}{2\Delta} f \rangle \ ,  \ \ \
V_{\pm z} =  {\bar V}_{\pm z} = 0 \ , \cr
\Pi_{yy} &=& g^{-1}- \langle
\frac{\zeta^2-\omega^2}{4\Delta^2} f \rangle \ ,  \ \ \
\Pi_{zz} = - \Pi_{yy} +  \langle f \rangle \ .
\label{result2}
\end{eqnarray}
In the presence of ``pinning'' frequency, $\Pi_{yy}$ should
be modified to
\begin{equation}
\Pi_{yy} = g^{-1} -  \langle
\frac{\zeta^2-\omega^2+\Omega^2_d}{4\Delta^2} f \rangle
\end{equation}
Now incorporating finite $\omega_L$ and using Eq.\ref{mixing}, we get the 
following results
\begin{eqnarray}
\chi^0_{+-} &=& \chi^0_{-+} = \left [ 
\langle \frac{\zeta^2+\Omega_L^2-\omega^2 f}{\zeta^2+\Omega_L^2-\omega^2}
 \rangle
-\frac{\langle f \rangle}{2}
-\frac{1}{2} \frac{\omega^2 \langle f \rangle^2}
{\langle (\omega^2-\zeta^2-\Omega^2_d) f \rangle} \right ] \ , \cr
\chi^0_{++} &=& \chi^0_{--} =  \left [ \frac{\langle f \rangle}{2}
-\frac{1}{2} \frac{\omega^2 \langle f \rangle^2}
{\langle (\omega^2-\zeta^2-\Omega^2_d) f \rangle } \right ] \ .
\end{eqnarray}
When $\omega \gg v_F q$, the transeverse
susceptibilities irreducible to $I$ in the long wavelength
limit ($q \ll k_F$) become
\begin{eqnarray}
\chi^0_{+-} &=& \chi^0_{+-} \approx N(0) \left [
-{1 \over 2} \frac{(v_F q)^2}{\omega^2} +
\frac{\omega^2 f_d \Omega_d^2}{ ( {1 \over 2} (v_F q)^2 -\omega^2 )
(\omega^2-{1 \over 2} (v_F q)^2 -\Omega_d^2) }
- {f_d \over 2} + {1 \over 2} {\omega^2 f_d \over \omega^2 - \Omega_d^2 -
{1 \over 2}(v_F q)^2} \right ] \ , \cr
\chi^0_{++} &=& \chi^0_{--} \approx N(0) \left [
 {f_d \over 2} - {1 \over 2} {\omega^2 f_d \over \omega^2 - \Omega_d^2 -
{1 \over 2}(v_F q)^2} \right ] \ .
\end{eqnarray}
Using the above results and the RPA (random phase approximation)
expression (see Eq.\ref{rpat}) for the full transverse susceptibility,
one finds two poles
which correspond to a propagating spin wave mode
and a damped mode. The dispersion relation of the
spin wave is the same as the one in the longitudinal
susceptibility:
\begin{equation}
\omega^2=(1-I f_d) \Omega_d^2 + \omega_L^2 
+\frac{1}{2}  f_d (v_F q)^2 \ ,
\end{equation}
where $\omega_L$ is the Larmor frequency.

\subsection{Magnetic field is parallel to
the order parameter: ${\bf H} \ || \ {\hat {\bf z}}$}

\subsubsection{Longitudinal susceptibility}
If the magnetic field is applied along the direction
of the order parameter, then the longitudinal susceptibility
corresponds to $\chi_{zz} \sim \langle [S_z,S_z] \rangle$.
As explained in previous sections, the full $\chi_{zz}$ can
be again obtained from
\begin{eqnarray}
\chi_{zz} &=& {\chi^0_{zz} \over 1 - I \chi^0_{zz}} \ , \cr
\chi^0_{zz} &=& \chi^{00}_{zz} +
{V_{zz} g {\bar V}_{zz} \over 1 - g \Pi_{zz}} \ .
\end{eqnarray}
Here $\chi^0_{zz}$ is the longitudinal susceptibility irreducible
with respect to $I$.
Notice that $\chi^{00}_{zz} \sim \langle [S_z,S_z] \rangle_0$,
$V_{zz} \sim \langle [S_z, \delta \Delta_z] \rangle_0,
{\bar V}_{zz} \sim \langle [\delta \Delta_z, S_z] \rangle_0$, and
$\Pi_{zz} \sim \langle [\delta \Delta_z, \delta \Delta_z] \rangle_0$,
where $\delta \Delta_z \sim
\Delta {\rm Tr}[\Psi^{\dagger} \alpha_3 \rho_1 \sigma_1 \Psi]$.

In this geometry the superconducting order parameter does not
move in the presence of the a.c. field.
So there is no coupling between the spin density and the 
fluctuation of the order parameter. As a result, we have
\begin{equation}
\chi_{zz} = {\chi^0_{zz} \over 1 - I \chi^0_{zz}} \ , 
\label{rpal}
\end{equation}
where
\begin{equation}
\chi^0_{zz} = \chi^{00}_{zz}=
\langle \frac{\zeta^2(1-f)}{\zeta^2-\omega^2} \rangle
\label{result3}
\end{equation}

Eq. \ref{rpal} gives a pole which gives the damped mode at 
\begin{equation}
\omega=i \frac{1}{\sqrt{2}} v_F |q| \sqrt{I (1-f_d)} \ .
\label{damp2}
\end{equation}

\subsubsection{Transverse susceptibility}

The transverse susceptibility,
$\chi_{+-} \sim {1 \over 2} \langle [S_x+iS_y,S_x-iS_y]
\rangle$, can be computed by using RPA expression (Eq.\ref{rpat})
and the following relation
\begin{eqnarray}
\chi^0_{+-} &=& \chi^{00}_{+-} +
\sum_{k=x,y} {V_{+k} g {\bar V}_{k-} \over 1 - g \Pi_{kk}} \ , \cr
\chi^0_{++} &=& \chi^{00}_{++} +
\sum_{k=x,y} {V_{+k} g {\bar V}_{k+} \over 1 - g \Pi_{kk}} \ ,
\label{mixing2}
\end{eqnarray}
Here $\chi^{00}_{+-} \sim \langle [S_x+iS_y,S_x-iS_y] \rangle_0$,
$\chi^{00}_{++} \sim \langle [S_x+iS_y,S_x+iS_y] \rangle_0$.
Also $V_{\pm k} \sim \langle [S_x \pm i S_y,\delta \Delta_k] \rangle_0$,
${\bar V}_{k \pm} \sim \langle [\delta \Delta_k,S_x \pm i S_y] \rangle_0$
and $\Pi_{kk} \sim
\langle [\delta \Delta_k,\delta \Delta_k] \rangle_0$
are spin/order-parameter couplings and the fluctuation
of the order parameter respectively.
In Appendix B, we show that $\chi^{00}_{+-}$, $V_{+-}$, 
and $\Pi_{+-}$ are the same as $\chi^0_{xx}, V_{xx}$, 
and $\Pi_{xx}$ when the magnetic field is along 
${\hat {\bf x}}$ direction. It is also shown that 
$\chi_{\pm\pm}=V_{\pm\pm}=\Pi_{\pm\pm}=0$. 
Thus we can use the previous results, Eq.\ref{result1} to get  
\begin{equation}
\chi^0_{+-} = \langle \frac{1}{\zeta^2-\omega^2} \left( \zeta^2
+\frac{\omega^2 \Omega_d^2 f}{\omega^2-\zeta^2-\Omega_d^2} \right)
\rangle
\label{result4}
\end{equation}
This is again exactly the same as $\chi^0_{xx}$, Eq.\ref{longx},  when the
magnetic field is along ${\hat {\bf x}}$ direction.
Since $\chi^0_{++} = \chi^0_{--} = 0$, the full susceptibility
is just given by
\begin{equation}
\chi_{+-} = {\chi^0_{+-} \over 1 - I \chi^0_{+-}} \ .
\end{equation}
This is also the same as the full longitudinal susceptibility
$\chi_{xx}$ for the case of ${\bf H} \ || \ {\hat {\bf x}}$.
\cite{note2}
Therefore, $\chi_{+-}$ in this case, supports a
propagating spin wave mode (there is no damped mode) and
the dispersion relation is given by
in Eq.\ref{mode2}.
The transverse response $\chi_{+-}$ when ${\bf H} \ || \
{\hat {\bf z}}$ is exactly the same as the longitudinal response
for the case of ${\bf H} \ || \ {\hat {\bf x}}$.

\section{Conclusion}

In this paper, we study dynamical spin susceptibilities
in a two dimensional $p$-wave superconductor with 
${\vec \Delta} = {\hat d} ({\hat k}_1 \pm i {\hat k}_2)$ symmetry. 
This order parameter vector has been a strong candidate for the
pairing symmetry of the superconducting Sr$_2$RuO$_4$.
Due to the fact that the direction of the order parameter vector
is fixed along a crystallographic direction, there are 
four possible susceptibility measurements: Longitudinal and 
transverse responses in the cases of the magnetic field 
parallel and perpendicular to the order parameter vector.
The existence of spin wave modes in each case is examined
and the dispersion relation is obtained. 
We found the following three modes.
Two of them are spin waves.

a) $\omega_{\parallel}^2=(1-I f_d) \Omega_d^2 
+\frac{1}{2} f_d (v_F q)^2$ \ ,

b) $\omega_{\perp}^2=\omega_{\parallel}^2+\omega_L^2$ \ ,

c) $\omega= i \frac{1}{\sqrt{2}} v_F |q| \sqrt{I (1-f_d)}$ \ .

The most crucial parameter is the pinning frequency.
It is most likely that the spin wave is observable for electron spin
resonance if $\Omega_d \ll \Delta (T)$.
We also believe that the experimental determination of $\Omega_d(T)$ will
provide an important insight in the pinning mechanism of ${\hat d}$.
Naturally this will provide another test of $p$-wave superconductivity.

As it is pointed out elsewhere\cite{maki3}, the superconductivity
in Bechgard like (TMTSF)$_2$PF$_6$, (TMTSF)$_2$ClO$_4$ etc. is most
likely of $p$-wave character as well.
Therefore it is highly desirable to look for the spin wave in the above
compounds as well.

Unlike in superfluid $^3$He, Sr$_2$RuO$_4$ is most likely in the vortex state.
Otherwise the magnetic field would penetrate only in the surface.
Nevertheless we believe that we can see the same expression for the spin
wave dispersion, if we reinterpret the superfluid density with the one in 
the vortex state.
We shall postpone the study of the spin wave in the vortex state 
to the future.

\acknowledgements
H.-Y.K. thanks S. Chakravarty and M. Sigrist for helpful discussions.
The work of H.-Y.K. was conducted under the auspices of the Department
of Energy, supported (in part) by funds provided by the University of
California for the conduct of discretionary research by Los Alamos
National Laboratory.
This work was also supported by Sloan Foundation Fellowship 
and the NSF CAREER Award No. DMR-9983731 (Y.B.K.),
and NSF grant DMR 95-31720 (K.M.). 
K.M. thanks Peter Fulde and Max-Planck Institut fur Physik Komplexer
Systeme at Dresden for their hospitality, where a part of this work
was carried out.

\appendix
\section{}
\subsection{Magnetic field is perpendicular to
the order parameter: ${\bf H} \ || \ {\hat {\bf x}}$}

\subsubsection{Longitudinal susceptibility}

Each correlation function can be computed from
\begin{eqnarray}
\chi^{00}_{xx} (i\omega_{\nu},{\bf q}) &=& T \sum_n 
\sum_{\bf p} {\rm Tr} [\alpha_1 G({\bf p},\omega_n)
\alpha_1 G({\bf p}-{\bf q},i\omega_n-i\omega_{\nu})] \ , \cr
V_{xx} (i\omega_{\nu},{\bf q}) &=& T \sum_n 
\sum_{\bf p} {\rm Tr} [\alpha_1 G({\bf p},\omega_n)
\alpha_1 \rho_1 \sigma_1 G({\bf p}-{\bf q},i\omega_n-i\omega_{\nu})] \ , \cr
\Pi_{xx} (i\omega_{\nu},{\bf q}) &=& T \sum_n 
\sum_{\bf p} {\rm Tr} [\alpha_1 \rho_1 \sigma_1 G({\bf p},\omega_n)
\alpha_1 \rho_1 \sigma_1 G({\bf p}-{\bf q},i\omega_n-i\omega_{\nu})] \ .
\end{eqnarray}

Summing over {\bf p} (circular Fermi surface is assumed) leads to
\begin{eqnarray}
\chi^{00}_{xx} (i\omega_{\nu}, {\bf q})
&=& \pi T N(0) \sum_n \left(1- \frac{\omega_n \omega_{n+\nu}+\Delta^2}
{\sqrt{\omega_n^2+\Delta^2} \sqrt{\omega_{n+\nu}^2+\Delta^2}} \right)
\frac{\sqrt{\omega_n^2+\Delta^2}+\sqrt{\omega_{n+\nu}^2+\Delta^2}}
{\left(\sqrt{\omega_n^2+\Delta^2}+\sqrt{\omega_{n+\nu}^2+\Delta^2}\right)^2
+\zeta^2} \ , \cr
V_{xx} (i\omega_{\nu}, {\bf q})
&=& \pi T N(0) \sum_n \left( \frac{ -i \omega_{\nu}\Delta}
{\sqrt{\omega_n^2+\Delta^2} \sqrt{\omega_{n+\nu}^2+\Delta^2}} \right)
\frac{\sqrt{\omega_n^2+\Delta^2}+\sqrt{\omega_{n+\nu}^2+\Delta^2}}
{\left(\sqrt{\omega_n^2+\Delta^2}+\sqrt{\omega_{n+\nu}^2+\Delta^2}\right)^2
+\zeta^2} \ , \cr
\Pi_{xx} (i\omega_{\nu}, {\bf q})
&=& \pi T N(0) \sum_n \left(1+ \frac{\omega_n \omega_{n+\nu}+\Delta^2}
{\sqrt{\omega_n^2+\Delta^2} \sqrt{\omega_{n+\nu}^2+\Delta^2}} \right)
\frac{\sqrt{\omega_n^2+\Delta^2}+\sqrt{\omega_{n+\nu}^2+\Delta^2}}
{\left(\sqrt{\omega_n^2+\Delta^2}+\sqrt{\omega_{n+\nu}^2+\Delta^2}\right)^2
+\zeta^2} \ ,
\label{complex}
\end{eqnarray}
where $\zeta={\bf v}_F \cdot {\bf q}$ and $N(0)=m/2\pi$ is 
the two dimensional density of states.
After summing over $\omega_n$ and analytic continuation 
$i \omega_{\nu} \rightarrow \omega + i \delta$, we get 
Eq.\ref{result1}.
 
\subsubsection{Transverse susceptibility}

Here $\chi^{00}_{+-} \sim \langle [S_y+iS_z,S_y-iS_z] \rangle_0$,
$\chi^{00}_{++} \sim \langle [S_y+iS_z,S_y+iS_z] \rangle_0$
are the bare susceptibilities and they can be computed from
\begin{eqnarray}
\chi^{00}_{+-} (i\omega_{\nu},{\bf q}) &=& T \sum_n 
\sum_{\bf p} {\rm Tr} [\alpha_{+} G({\bf p},\omega_n)
\alpha_{-} G({\bf p}-{\bf q},i\omega_n-i\omega_{\nu})] \ , \cr
\chi^{00}_{++} (i\omega_{\nu},{\bf q}) &=& T \sum_n 
\sum_{\bf p} {\rm Tr} [\alpha_{+} G({\bf p},\omega_n)
\alpha_{+} G({\bf p}-{\bf q},i\omega_n-i\omega_{\nu})] \ ,
\end{eqnarray}
where $\alpha_{\pm} = 
{1 \over \sqrt{2}}(\alpha_2 \pm i \alpha_3)$ respresenting the
transverse spin vertices: ${1 \over \sqrt{2}}(S_y \pm i S_z)
\sim {\rm Tr}[\Psi^{\dagger} \alpha_{\pm} \Psi]$.
Similarly, we have 
\begin{eqnarray}
V_{\pm k} (i\omega_{\nu},{\bf q}) &=& T \sum_n 
\sum_{\bf p} {\rm Tr} [\alpha_{\pm} G({\bf p},\omega_n)
\alpha_k \rho_1 \sigma_1 G({\bf p}-{\bf q},i\omega_n-i\omega_{\nu})] \ , \cr
\Pi_{kk} (i\omega_{\nu},{\bf q}) &=& T \sum_n 
\sum_{\bf p} {\rm Tr} [\alpha_k \rho_1 \sigma_1 G({\bf p},\omega_n)
\alpha_k \rho_1 \sigma_1 G({\bf p}-{\bf q},i\omega_n-i\omega_{\nu})] \ .
\end{eqnarray}
It is also useful to represent these correlation functions in terms of 
$V_{(\pm, \pm)} \sim \langle [S_y \pm i S_z,\delta 
\Delta_{\pm}] \rangle_0$ and  
$\Pi_{(\pm, \pm)} \sim \langle [\delta \Delta_{\pm},\delta \Delta_{\pm}] 
\rangle_0$, where $\delta \Delta_{\pm} \sim {\rm Tr}[\Psi^{\dagger} 
\alpha_{\pm} \rho_1 \sigma_1 \Psi]$. It is found that
\begin{eqnarray}
V_{\pm y} &=& {\bar V}_{y \pm} = \frac{1}{\sqrt{2}} 
(V_{\pm \pm} + V_{\pm \mp}) \ , \cr
V_{\pm z} &=& {\bar V}_{z \pm} = \frac{1}{\sqrt{2}i} 
(\pm V_{\pm \pm} \mp V_{\pm \mp}) \ , \cr
\Pi_{yy} &=& \frac{1}{2} (\Pi_{++}+\Pi_{+-}+\Pi_{-+}+\Pi_{--}) \ , \cr 
\Pi_{zz} &=& - \frac{1}{2} (\Pi_{++}-\Pi_{+-}-\Pi_{-+}+\Pi_{--}) \ ,
\end{eqnarray}
where 
\begin{eqnarray}
V_{(\pm, \pm)} (i\omega_{\nu},{\bf q}) &=& T \sum_n 
\sum_{\bf p} {\rm Tr} [\alpha_{\pm} G({\bf p},\omega_n)
\alpha_{\pm} \rho_1 \sigma_1 
G({\bf p}-{\bf q},i\omega_n-i\omega_{\nu})] \ , \cr
\Pi_{(\pm,\pm)} (i\omega_{\nu},{\bf q}) &=& T \sum_n 
\sum_{\bf p} {\rm Tr} [\alpha_{\pm} \rho_1 \sigma_1 G({\bf p},\omega_n)
\alpha_{\pm} \rho_1 \sigma_1 G({\bf p}-{\bf q},i\omega_n-i\omega_{\nu})] \ .
\end{eqnarray}

After summing over {\bf p}, we get
\begin{eqnarray}
\chi^{00}_{+-} &=& \chi^{00}_{-+}=
\pi T N(0) \sum_n \left(1- \frac{\omega_n \omega_{n+\nu}}
{\sqrt{\omega_n^2+\Delta^2} \sqrt{\omega_{n+\nu}^2+\Delta^2}} \right)
\frac{\sqrt{\omega_n^2+\Delta^2}+\sqrt{\omega_{n+\nu}^2+\Delta^2}}
{\left(\sqrt{\omega_n^2+\Delta^2}+\sqrt{\omega_{n+\nu}^2+\Delta^2}\right)^2
+\zeta^2} \ , \cr
\chi^{00}_{++} &=&
\chi^{00}_{--} =
\pi T N(0) \sum_n \left(\frac{-\Delta^2}
{\sqrt{\omega_n^2+\Delta^2} \sqrt{\omega_{n+\nu}^2+\Delta^2}} \right)
\frac{\sqrt{\omega_n^2+\Delta^2}+\sqrt{\omega_{n+\nu}^2+\Delta^2}}
{\left(\sqrt{\omega_n^2+\Delta^2}+\sqrt{\omega_{n+\nu}^2+\Delta^2}\right)^2
+\zeta^2} \ , \cr
V_{+-} &=& V_{-+} = V_{++} = V_{--}
= \pi T N(0) \sum_n \left(\frac{1}{2} \frac{- i \omega_{\nu}\Delta}
{\sqrt{\omega_n^2+\Delta^2} \sqrt{\omega_{n+\nu}^2+\Delta^2}} \right)
\frac{\sqrt{\omega_n^2+\Delta^2}+\sqrt{\omega_{n+\nu}^2+\Delta^2}}
{\left(\sqrt{\omega_n^2+\Delta^2}+\sqrt{\omega_{n+\nu}^2+\Delta^2}\right)^2
+\zeta^2} \ , \cr
\Pi_{+-} &=& \Pi_{-+} =
\pi T N(0) \sum_n \left(\frac{\Delta^2}
{\sqrt{\omega_n^2+\Delta^2} \sqrt{\omega_{n+\nu}^2+\Delta^2}} \right)
\frac{\sqrt{\omega_n^2+\Delta^2}+\sqrt{\omega_{n+\nu}^2+\Delta^2}}
{\left(\sqrt{\omega_n^2+\Delta^2}+\sqrt{\omega_{n+\nu}^2+\Delta^2}\right)^2
+\zeta^2} \ , \cr
\Pi_{++} &=& \Pi_{--} =
\pi T N(0) \sum_n \left(1+ \frac{\omega_n \omega_{n+\nu}}
{\sqrt{\omega_n^2+\Delta^2} \sqrt{\omega_{n+\nu}^2+\Delta^2}} \right)
\frac{\sqrt{\omega_n^2+\Delta^2}+\sqrt{\omega_{n+\nu}^2+\Delta^2}}
{\left(\sqrt{\omega_n^2+\Delta^2}+\sqrt{\omega_{n+\nu}^2+\Delta^2}\right)^2
+\zeta^2} \ ,
\end{eqnarray}
Now summing over $\omega_n$ and analytic continuation 
$i \omega_{nu} \rightarrow \omega + i \delta$ lead to
Eq.\ref{result2}.

\section{}
\subsection{Magnetic field is parallel to
the order parameter: ${\bf H} \ || \ {\hat {\bf z}}$}

\subsubsection{Longitudinal susceptibility}

Notice that $\chi^{00}_{zz} \sim \langle [S_z,S_z] \rangle_0$,
$V_{zz} \sim \langle [S_z, \delta \Delta_z] \rangle_0,
{\bar V}_{zz} \sim \langle [\delta \Delta_z, S_z] \rangle_0$, and 
$\Pi_{zz} \sim \langle [\delta \Delta_z, \delta \Delta_z] \rangle_0$,
where $\delta \Delta_z \sim 
\Delta {\rm Tr}[\Psi^{\dagger} \alpha_3 \rho_1 \sigma_1 \Psi]$.
These correlations functions can be expressed as
\begin{eqnarray}
\chi^{00}_{zz} (i\omega_{\nu},{\bf q}) &=& T \sum_n 
\sum_{\bf p} {\rm Tr} [\alpha_3 G({\bf p},\omega_n)
\alpha_3 G({\bf p}-{\bf q},i\omega_n-i\omega_{\nu})] \ , \cr
V_{zz} (i\omega_{\nu},{\bf q}) &=& T \sum_n 
\sum_{\bf p} {\rm Tr} [\alpha_3 G({\bf p},\omega_n)
\alpha_3 \rho_1 \sigma_1 G({\bf p}-{\bf q},i\omega_n-i\omega_{\nu})] \ , \cr
\Pi_{zz} (i\omega_{\nu},{\bf q}) &=& T \sum_n 
\sum_{\bf p} {\rm Tr} [\alpha_3 \rho_1 \sigma_1 G({\bf p},\omega_n)
\alpha_3 \rho_1 \sigma_1 G({\bf p}-{\bf q},i\omega_n-i\omega_{\nu})] \ .
\end{eqnarray}
>From the consideration of the matrix elements, we can easily see that 
$V_{zz} = {\bar V}_{zz} = 0$. Therefore, there is no mixing between
spin fluctuations and the order paramter in this case and 
$\chi^0_{zz}$ is just given by $\chi^{00}_{zz}$.
After summing over {\bf p}, we get 
\begin{equation}
\chi^{00}_{zz} (i\omega_{\nu}, {\bf q})=
\pi T N(0) \sum_n \left(1- \frac{\omega_n \omega_{n+\nu}-\Delta^2}
{\sqrt{\omega_n^2+\Delta^2} \sqrt{\omega_{n+\nu}^2+\Delta^2}} \right)
\frac{\sqrt{\omega_n^2+\Delta^2}+\sqrt{\omega_{n+\nu}^2+\Delta^2}}
{\left(\sqrt{\omega_n^2+\Delta^2}+\sqrt{\omega_{n+\nu}^2+\Delta^2}\right)^2
+\zeta^2} \ .
\end{equation}
Summing over $\omega_n$ and analytic continuation lead to
Eq.\ref{result3}.
 
\subsubsection{Transverse susceptibility}
Here $\chi^{00}_{+-} \sim \langle [S_x+iS_y,S_x-iS_y] \rangle_0$,
$\chi^{00}_{++} \sim \langle [S_x+iS_y,S_x+iS_y] \rangle_0$.
Also $V_{\pm k} \sim \langle [S_x \pm i S_y,\delta \Delta_k] \rangle_0$,
${\bar V}_{k \pm} \sim \langle [\delta \Delta_k,S_x \pm i S_y] \rangle_0$,
and $\Pi_{kk} \sim 
\langle [\delta \Delta_k,\delta \Delta_k] \rangle_0$.
These correlation function can be again calculated from
\begin{eqnarray}
\chi^{00}_{+\mp} (i\omega_{\nu},{\bf q}) &=& T \sum_n 
\sum_{\bf p} {\rm Tr} [\beta_{+} G({\bf p},\omega_n)
\beta_{\mp} G({\bf p}-{\bf q},i\omega_n-i\omega_{\nu})] \ , \cr
V_{\pm k} (i\omega_{\nu},{\bf q}) &=& T \sum_n 
\sum_{\bf p} {\rm Tr} [\beta_{\pm} G({\bf p},\omega_n)
\alpha_k \rho_1 \sigma_1 G({\bf p}-{\bf q},i\omega_n-i\omega_{\nu})] \ , \cr
\Pi_{kk} (i\omega_{\nu},{\bf q}) &=& T \sum_n 
\sum_{\bf p} {\rm Tr} [\alpha_k \rho_1 \sigma_1 G({\bf p},\omega_n)
\alpha_k \rho_1 \sigma_1 G({\bf p}-{\bf q},i\omega_n-i\omega_{\nu})] \ ,
\end{eqnarray}
where $\beta_{\pm} = {1 \over \sqrt{2}}
(\alpha_1 \pm i \alpha_2)$ respresenting the
transverse spin vertices: ${1 \over \sqrt{2}}(S_x \pm i S_y)
\sim {\rm Tr}[\Psi^{\dagger} \beta_{\pm} \Psi]$.
One can also rewrite these correlation functions in terms of 
$V_{(\pm, \pm)} \sim \langle [S_x \pm i S_y,\delta \Delta_{\pm}] \rangle_0$ 
$\Pi_{(\pm, \pm)} \sim \langle [\delta \Delta_{\pm},\delta \Delta_{\pm}] 
\rangle_0$, where $\delta \Delta_{\pm} \sim {\rm Tr}[\Psi^{\dagger} 
\beta_{\pm} \rho_1 \sigma_1 \Psi]$:
\begin{eqnarray}
V_{\pm x} &=& {\bar V}_{y \pm} = \frac{1}{\sqrt{2}} 
(V_{\pm \pm} + V_{\pm \mp}) \ , \cr
V_{\pm y} &=& {\bar V}_{z \pm} = \frac{1}{\sqrt{2}i} 
(\pm V_{\pm \pm} \mp V_{\pm \mp}) \ , \cr
\Pi_{xx} &=& \frac{1}{2} (\Pi_{++}+\Pi_{+-}+\Pi_{-+}+\Pi_{--}) \ , \cr 
\Pi_{yy} &=& - \frac{1}{2} (\Pi_{++}-\Pi_{+-}-\Pi_{-+}+\Pi_{--}) \ ,
\end{eqnarray}
where 
\begin{eqnarray}
V_{(\pm, \pm)} (i\omega_{\nu},{\bf q}) &=& T \sum_n 
\sum_{\bf p} {\rm Tr} [\beta_{\pm} G({\bf p},\omega_n)
\beta_{\pm} \rho_1 \sigma_1 
G({\bf p}-{\bf q},i\omega_n-i\omega_{\nu})] \ , \cr
\Pi_{(\pm,\pm)} (i\omega_{\nu},{\bf q}) &=& T \sum_n 
\sum_{\bf p} {\rm Tr} [\beta_{\pm} \rho_1 \sigma_1 G({\bf p},\omega_n)
\beta_{\pm} \rho_1 \sigma_1 G({\bf p}-{\bf q},i\omega_n-i\omega_{\nu})] \ .
\end{eqnarray}

From the matrix elements, one can see that
\begin{equation}
\chi^{00}_{++}=\chi^{00}_{--}=0,  \ \ \ 
V_{++}=V_{--}=0,  \ \ \
\Pi_{++}=\Pi_{--}=0 \ .
\end{equation} 
Thus Eq.\ref{mixing2} can be simplified as
\begin{eqnarray}
\chi^0_{+-} &=& \chi^0_{-+} = \chi^{00}_{+-} + 
{V_{+-} g {\bar V}_{+-} \over 1 - g \Pi_{+-}} \ , \cr
\chi^0_{++} &=& \chi^0_{--} = 0 \ .  
\end{eqnarray}
After summing over {\bf p}, we obtain
the same equation as Eq. \ref{complex}.
Thus we can use the previous results (Eq.\ref{result1}) to get
Eq.\ref{result4} which is same as Eq.\ref{longx}.


\begin{references}

\bibitem{cava} R. J. Cava, B. Batlogg, K. Kiyano, H. Takagi,
J. J. Krajewski, W. F. Peck, L. W. Rupp, and  C. Chen,  
Phys. Rev. B {\bf 49}, 11890 (1994).
\bibitem{maeno1} Y. Maeno, H. Hashmoto, K. Yoshida, S. Nishizaki,
T. Fujita, J. G. Bednoz, and F. Lichtenberg,
Nature {\bf 372}, 532 (1994).
\bibitem{maeno1-1} Y. Maeno, S. Nishizaki, K. Yoshida, 
S. Ikeda, and T. Fujita,
J. Low Temp. Phys. {\bf 105}, 1577 (1996).
\bibitem{maeno2} Y. Maeno,  
Physica C {\bf 282}-{\bf 287}, 206 (1997).
\bibitem{mac1} A. P. Mackenzie, S. R. Julian, A. J. Diver,
G. G. Lonzarich, Y. Maeno, S. Nishizaki, and T. Fujita,
Phys. Rev. Letts. {\bf 76}, 3786 (1996).
\bibitem{band} T. Oguchi, 
Phys. Rev. B {\bf 51}, 1385 (1995);
D. J. Singh, Phys. Rev. B {\bf 52}, 1358 (1995).    
\bibitem{sigrist1} T. M. Rice and M. Sigrist, 
J. Phys. Condens. Matter {\bf 7}, L643 (1995).
\bibitem{baskaran} G. Baskaran, 
Physica B {\bf 223} \& {\bf 224}, 490 (1996).
\bibitem{mazin2} I. I. Mazin and D. J. Singh, 
Phys. Rev. Lett. {\bf 79}, 733 (1997).
\bibitem{mac2} A. P. Mackenzie, R. K. W. Haselwimmer, A. W. Tyler, 
G. G. Lonzarich, Y. Mori, S. Nishizaki,  and Y. Maeno,
Phys. Rev. Letts. {\bf 80}, 161 (1998).
\bibitem{maeno4} K. Ishida, Y. Kitaoka, K. Asayama, S. Ikeda,
S. Nishizaki, Y. Maeno, K. Yoshida,  and T. Fujita,
Phys. Rev. B {\bf 56}, R505 (1997).
\bibitem{liu} R. Jin, Yu Zadorozhny, Y. Liu, D. G. Schlom,
Y. Mori, and Y. Maeno, 
Phys. Rev. B {\bf 59}, 4433 (1999).
\bibitem{ishida} K. Ishida, H. Mukuda, Y. Kitaoka, K. Asayama, Z. Q. Mao,
Y. Mori, and  Y. Maeno,  Nature {\bf 396} 658 (1998).
\bibitem{vollhardt} {\it The Superfluid Phases of Helium 3}, 
D. Vollhardt and P. W\"olfle (Tayor \& Francis, New York, 1990).
\bibitem{nishizaki} S. Nishizaki, Y. Maeno, S. Farner, S. Ikeda, 
 and T. Fujita, 
 J. Phys. Soc. Jpn. {\bf 67} 560 (1998).
\bibitem{nishizaki2} S. Nishizaki, Z. Q. Mao, and Y. Maeno (private 
communication)  
\bibitem{sigrist2} M. Sigrist and M. E. Zhitomirsky, J. Phys. Soc. Jpn.
{\bf 67} 3452 (1996).
\bibitem{agt1} D. F. Agterberg, T. M. Rice, and M. Sigrist,  
Phys. Rev. Letts. {\bf 78}, 3374 (1997).
\bibitem{agt2} D. R. Agterberg, preprint.
\bibitem{maeno5} G. M. Luke, Y. Fudamoto, K. M. Kojima, M. I. Larkin,
J. Merrin, B. Nachumi, Y. J. Uemura, Y. Maeno, Z. Q. Mao,
Y. Mori, H. Nakamura, and M. Sigrist, 
Nature {\bf 394}, 558 (1998).
\bibitem{mota} A. C. Mota, E. Dumont, A. Amann, and Y. Maeno, Physica B,
 {\bf 259-261} 934 (1999).
\bibitem{heffner} R. H. Heffner and M. R. Norman, Comments on Condensed Matter
Physics {\bf 17} 361 (1996).
\bibitem{mac3} T. M. Riseman, P. G. Kealey, E. M. Forgan, 
A. P. Mackenzie, L. M. Galvin, A. W. Tyler, S. L. Lee, 
C. Ager, D. McPaul, C. M. Agterberg, R. Cubitt, Z. Q. Mao, 
S. Akima, and Y. Maeno, 
Nature {\bf 396}, 242 (1998).
\bibitem{agt3} D. F. Agterberg,
Phys. Rev. Letts. {\bf 80}, 5184 (1998).
\bibitem{agt4} D. F. Agterberg and R. Heeb, Phys. Rev. B {\bf 59}
7076 (1999).
\bibitem{wang} G. F. Wang and K. Maki, Europhys. Lett. {\bf 45} 71 (1999)
\bibitem{maki} K. Maki and E. Puchkaryov, Europhys. Lett. {\bf 45} 263 (1999).
\bibitem{matsumoto} M. Matsumoto and M. Sigrist, J. Phys. Soc. Jpn.
{\bf 68} 724 (1999).
\bibitem{tewordt} L. Tewordt, Phys. Rev. Lett., {\bf 83} 1007 (1999).
\bibitem{maki2} K. Maki and H. Ebisawa, 
Prog. Theor. Phys. (Kyoto), {\bf 50}, 1452 (1973); {\bf 51} 690 (1974).
\bibitem{note} Here we have used $(v_F q)^2$ dependent term
from Ref. [14]. 
To get the correct $(v_F q)^2$ dependent term, the  more consistent
treatment is required. [See for example A. Virosztek and K. Maki,
Phys. Rev. B {\bf 48} 1368 (1993)]  
\bibitem{note2} We assume that $\omega_L^2 << \Omega_d^2$.
\bibitem{maki3} K. Maki, H. Won, M. Kohmoto, J. Shiraishi, Y. Morita, 
and G. F. Wang, Physica C {\bf 317-318} 353 (1999).



\end{references}
\end{document}